\documentclass[prb,twocolumn,showpacs,superscriptaddress]{revtex4}

\usepackage{graphicx}
\usepackage{dcolumn}
\usepackage{amsmath}
\usepackage{color}

\newcommand{\SrCrO}{Sr$_{3}$Cr$_2$O$_8$}
\newcommand{\BaCrO}{Ba$_{3}$Cr$_2$O$_8$}

\begin{document}
\title{Orbital fluctuations and orbital order below the Jahn-Teller transition in Sr$_{3}$Cr$_2$O$_8$}

\author{Zhe~Wang}
\author{M.~Schmidt}
\author{A.~G\"{u}nther}
\author{S.~Schaile}
\author{N.~Pascher}
\author{F.~Mayr}
\affiliation{Experimental Physics V, Center for Electronic
Correlations and Magnetism, Institute of Physics, University of Augsburg, D-86135 Augsburg, Germany}

\author{Y.~Goncharov}
\affiliation{Experimental Physics V, Center for Electronic
Correlations and Magnetism, Institute of Physics, University of Augsburg, D-86135 Augsburg, Germany}
\affiliation{Institute of General Physics, Russian Academy of Sciences, 119991 Moscow, Russia}

\author{D. L. Quintero-Castro}
\affiliation{Helmholtz-Zentrum Berlin f\"{u}r Materialien und
Energie, D-14109 Berlin, Germany}

\affiliation{Institut f\"{u}r Festk\"{o}rperphysik, Technische
Universit\"{a}t Berlin, D-10623 Berlin, Germany}
\author{A. T. M. N. Islam}

\affiliation{Helmholtz-Zentrum Berlin f\"{u}r Materialien und
Energie, D-14109 Berlin, Germany}

\author{B. Lake}
\affiliation{Helmholtz-Zentrum Berlin f\"{u}r Materialien und
Energie, D-14109 Berlin, Germany}

\affiliation{Institut f\"{u}r Festk\"{o}rperphysik, Technische
Universit\"{a}t Berlin, D-10623 Berlin, Germany}

\author{H.-A.~Krug von Nidda}
\author{A.~Loidl}
\author{J.~Deisenhofer}
\affiliation{Experimental Physics V, Center for Electronic
Correlations and Magnetism, Institute of Physics, University of Augsburg, D-86135 Augsburg, Germany}

\date{\today}

\begin{abstract}
We report on the magnetic and phononic excitation spectrum of
\SrCrO~determined by THz and infrared (IR) spectroscopy, and
electron spin resonance (ESR) measurements across the Jahn-Teller
transition, which is detected by specific-heat measurements
to occur at $T_{JT}$ = 285~K. We identify the singlet-triplet
excitations in the dimerized ground state and estimate the exchange
couplings in the system. ESR absorptions were observed up
to $T^{*}$ = 120~K with a linewidth
\begin{math}\propto \exp{(-\Delta/k_BT)}\end{math} and $\Delta/k_B$ = 388~K indicating a phonon-mediated spin relaxation via the excited
orbital state of the Cr $e$ doublet in the orbitally ordered state.
Upon entering the low-symmetry Jahn-Teller distorted phase below
$T_{JT}$, we find an extended regime $T^{*}<T<T_{JT}$
where the IR active phonons change only gradually with decreasing
temperature. This regime is associated with strong fluctuations in
the orbital and lattice degrees of freedom in agreement with the
loss of the ESR signal above $T^{*}$.  Using the measured magnetic
and phononic excitation spectrum we model the orbital contribution
to the specific heat and find the persistence of strong fluctuations
far below $T_{JT}$.
\end{abstract}

% 78.40.-q    Absorption and reflection spectra: visible and ultraviolet (for infrared spectra, see 78.30.-j)
% 78.20.-e    Optical properties of bulk materials and thin films (for optical properties related to materials treatment,
%             see 81.40.Tv; for optical materials, see 42.70-a; for optical properties of superconductors,
%             see 74.25.Gs; for optical properties of rocks and minerals, see 91.60.Mk)
% 71.70.-d    Level splitting and interactions (see also 73.20.-r Surface and interface electron states;
% 71.70.Ch    Crystal and ligand fields
% 71.70.Ej    Spin-orbit coupling, Zeeman and Stark splitting, Jahn-Teller effect
% 75.30.Fv    Spin-density waves
% 75.20.Hr    Local moment in compounds and alloys; Kondo effect, valence fluctuations, heavy fermions
% 71.70.Ch    Crystal and ligand fields
% 76.30.-v    Electron paramagnetic resonance and relaxation

\pacs{78.30.-j,71.70.-d,76.30.-v,78.20.-e}

\maketitle

%\section{Introduction}

Orbital degrees of freedom (DOF) and their coupling to other DOF
play an important role in understanding the physics in
many transition-metal compounds including colossal magnetoresistive
manganites, vanadates, and iron-based
superconductors.\cite{Tokura,Khaliullin05,Krueger09} Orbital
ordering (OO) mechanisms,\cite{Kugel82} collective orbital
excitations,\cite{Saitoh01} and frustration effects in the orbital
sector have attracted considerable attention and formed the research
field of \textit{orbital physics}. Exotic ground states such as
orbital and spin-orbital liquids have been explored both
experimentally and theoretically.\cite{Khaliullin00} Orbital
fluctuations play an important role in the formation of these states
and may even induce a dimerization of spins via magnetoelastic
coupling.\cite{vandenBrink11}

Here we investigate the OO transition in the spin-gapped dimerized
system \SrCrO. This compound has come into focus due to the
occurrence of a field-induced magnon condensation.\cite{Giamarchi08,Aczel09}
The room
temperature crystal structure of \SrCrO ~was determined to be
hexagonal with space group \emph{R}$\bar{3}$\emph{m}.\cite{Cuno89}
Each Cr$^{5+}$ ion with spin $S$ = 1/2 is surrounded by an oxygen
tetrahedron and couples antiferromagnetically to adjacent Cr$^{5+}$
ions along \emph{c} direction, forming a spin singlet ground
state at low temperatures. The tetrahedral crystal field splits the
3\emph{d} levels into lower-lying doubly-degenerated \emph{e} and
triply-degenerated \emph{t} orbitals. Thus, the Cr$^{5+}$ ions are
Jahn-Teller (JT) active and a JT transition to a monoclinic
structure with three twinned domains and space group \emph{C}2/\emph{c} has been reported by
neutron experiments to occur at 275 K.\cite{Chapon08,Castro10}
Primarily associated with the antiferrodistortive displacement of
the apical oxygen ions this transition is suggested to be
accompanied by the splitting of the \emph{e} doublet into a lower-lying $d_{3z^2-r^2}$ and excited $d_{x^2-y^2}$ orbital, and an
antiferro-orbital ordering of $d_{3z^2-r^2}$ orbitals.  In addition,
this ordering reportedly changes the magnetic exchange paths and
leads to spatially anisotropic exchange couplings.\cite{Chapon08,Castro10}
Recent \emph{ab initio} calculations
showed that correlation effects within Cr-3\emph{d} orbitals can
account for both the structural phase transition and the singlet
ground state.\cite{Radtke10}

We present a combined study using THz and infrared
(IR) spectroscopy, electron spin resonance (ESR) and specific heat.
We directly observe the singlet-triplet excitations in the ground
state and derive the splitting of the $e$-orbitals from the ESR spin
relaxation below 120~K. Above 120~K strong fluctuations
are found to influence both the IR active lattice vibrations and the
spin relaxation up to the JT transition at 285~K.

%The orbital contribution to the specific heat is extracted by using
%the obtained magnetic and phononic excitation spectrum of \SrCrO~
%and is consistent with an extended orbital fluctuation regime below
%the JT transition.

%Orbital fluctuation regimes as a precursor above Jahn-Teller driven
%orbital ordering transitions have been reported previously
%\cite{Zhou07,Tsurkan10}, but the present case marks

%Using THz spectroscopy in magnetic fields we determine the
%singlet-triplet transitions in the dimerized ground state and the
%exchange coupling in the system. The infrared active phonons tracked
%as a function of temperature show that the reported structural and
%orbital-order transition at about 275~K is not complete but
%significant fluctuations remain down to about 125~K. Using the
%experimentally observed excitation spectrum we are able to model the
%magnetic and lattice contributions to the specific heat of the
%system and reveal the entropy associated with the orbital degrees of
%freedom in the system. The ESR absorption lines broaden
%exponentially with temperature and become untraceable above 125~K
%indicating the correlation between spin relaxation and lattice
%fluctuations.

%{\color{red}The eigen-equations of the singlet and triplet states are given in Appendix A.}

%\section{Experimental details}\label{Sec:ExperDetail}
Single crystals grown by the floating-zone method\cite{Castro10,Islam10} were
orientated by Laue diffraction and cut along an
$a_hc_h$-plane, where the subscript $h$ stands for hexagonal and a subscript $m$ for monoclinic in the following.
Heat capacity was measured in a Quantum Design
physical properties measurement system from 1.8 to 300~K. Susceptibility was measured using a
SQUID magnetometer (Quantum Design). Polarization-dependent
reflectivity was measured from 20 to 300~K in the
far- and mid-IR range using the Bruker Fourier-transform IR
spectrometers IFS 113v and IFS 66v/S with a He-flow cryostat
(Cryovac). THz transmission experiments were performed in Voigt
configuration using a Mach-Zehnder-type interferometer with
backward-wave oscillators covering the frequency range 115~GHz -
1.4~THz and a magneto-optical cryostat (Oxford/Spectromag) with
applied magnetic fields up to 7~T. ESR measurements were performed
in a Bruker ELEXSYS E500 CW-spectrometer at X-band frequency of 9.48~GHz from 4 to 300~K.

%\section{Experimental results and discussion}
%
%\subsection{THz spectroscopy: singlet-triplet excitations in the dimerized state}\label{Sec:ThzTrans}

\begin{figure}[t]
\centering
\includegraphics[width=85mm,clip]{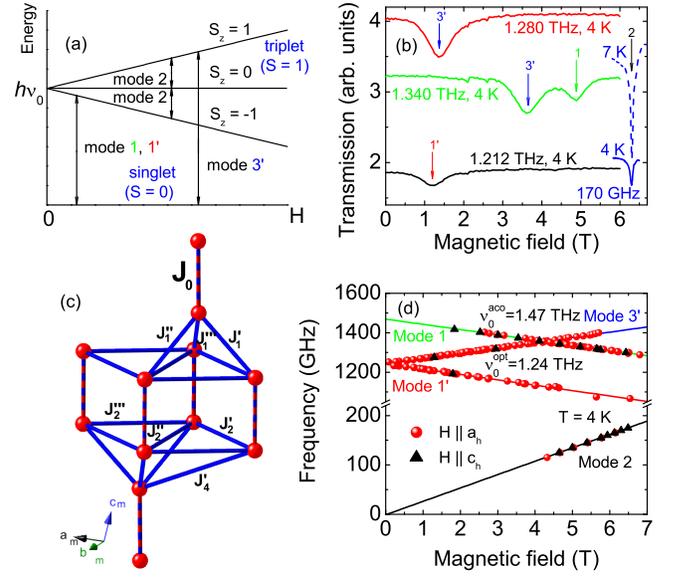}
\vspace{2mm} \caption[]{\label{Fig:EnergyScheme} (Color online) (a)
Zeeman-splitting of triplet states in a magnetic field. Modes 1,
1', 2, and 3' are described in the text. (b) Transmission spectra
measured at different frequencies at 4~K. The spectrum obtained at 170 GHz and 7~K (dashed line) is shifted for clarity. The spectra corresponding to mode 2 are measured with \textbf{H} $\|$ $c_{h}$, while the spectra corresponding to modes 1, 1' and 3' are measured with \textbf{H} $\|$ $a_{h}$. (c) Bilayer structure of Cr$^{5+}$ ions. (d) Magnetic
field dependence of the observed absorption frequencies at 4~K for \textbf{H} $\|$ $c_{h}$ and \textbf{H} $\|$ $a_{h}$.}
\end{figure}

The results of all THz transmission spectra measured with different
frequencies for \textbf{H} $\|$ $c_{h}$ and \textbf{H} $\|$ $a_{h}$
are summarized in Fig.~\ref{Fig:EnergyScheme}(d). Absorptions
labeled 1, 1', and 3' in Fig.~\ref{Fig:EnergyScheme}(b) correspond
to excitations from the singlet ground state to the excited triplet
states Zeeman-split by the external magnetic field \textbf{H}, while
absorption 2 corresponds to the intra-triplet excitations as
illustrated in Fig. \ref{Fig:EnergyScheme}(a).  The excitation
spectrum agrees with the one reported for \BaCrO~by Kofu \textit{et
al.}\cite{Kofu09b} except for mode 3' not having been observed for
\BaCrO. Modes 3' and 1' can be described in terms of a linear Zeeman
splitting following $h\nu =h\nu _{0}\pm g\mu_{B}H$ with
$\nu_{0}^{opt}=1.24$~THz~$\cong5.13$~meV and an effective
\emph{g}-factor of about 1.92(3) in agreement with the value of 1.94
of mode 2, with ESR data discussed below and reported
values for \BaCrO~and \SrCrO.\cite{Kofu09b,Aczel09} For mode 1 we
could observe the lower branch within the available frequency range
and found $\nu_{0}^{aco}=1.47$~THz~$\cong6.08$~meV which corresponds
well to the magnetic excitation energy reported by neutron
scattering at the $\Gamma$-point.\cite{Castro10} Following
Ref.~\onlinecite{Kofu09b} we assign modes 1 and 1' to the cooperative
acoustic ($n=0$) and optical modes ($n=\pm1$) of the coupled dimers
of different bilayers with eigenfrequencies
$h\nu_{0}\simeq\sqrt{J_{0}^{2}+J_{0}\gamma}$ where
$\gamma=2[(J'_{1}+J''_{1}+J'''_{1})\cos(\frac{2}{3}n\pi)+(J'_{2}+J''_{2}+J'''_{2})+(J'_{4}+J''_{4}+J'''_{4})\cos(\frac{2}{3}n\pi)]$.
Here $J_{0}$ denotes the intradimer interaction, $J_{1}$s, $J_{2}$s,
and $J_{4}$s are the interdimer interactions (see
Fig.~\ref{Fig:EnergyScheme}(c)). Using the values of exchange
couplings determined from neutron scattering \cite{Castro10} we find
$h\nu_{0}^{aco}=5.92$~meV and $h\nu_{0}^{opt}=5.14$~meV in
agreement with our experimental values. The observability of the
singlet-triplet transitions 1, 1', and 3' implies an additional
anisotropic contribution to the spin Hamiltonian, which mixes
singlet and triplet states and relaxes the selection rule $\Delta S$.\cite{Glazkov04}
The intensity of mode 2
increases with increasing temperature in agreement with the thermal
population of the excited levels (see absorption lines at 4~K and
7~K in Fig.~\ref{Fig:EnergyScheme}(b)) and corresponds to the
expected ESR signal which has been tracked with high sensitivity in
a cavity-based setup at X-band frequency.

%Figure 1 Singlet & Triplet transition

% i.e. the external magnetic field $H ~\|$ \emph{a,b}-plane or $H ~\bot$ \emph{a,b}-plane and the electric field of incident wave $E ~\|~$\emph{a,b}-plane or $E ~\bot$ \emph{a,b}-plane.

%\subsection{ESR: spin-lattice relaxation}

%\emph{An X-band ESR absorption spectrum measured at 30~K is shown in
%Fig.~\ref{Fig:ESR}(a), which is well fitted by a Lorentzian line
%shape characterized by the effective $g$ factor, its half-width
%half-maximum linewidth $\Delta H$, and its double integrated
%intensity $I_{ESR}$.}

\begin{figure}[htb]
\centering
\includegraphics[width=70mm,clip]{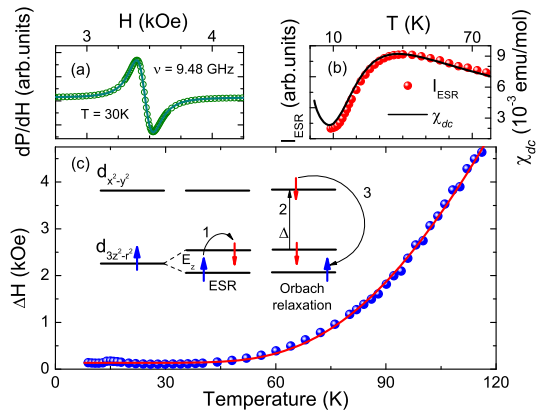}
\vspace{2mm} \caption[]{\label{Fig:ESR} (Color online) (a) X-band
ESR spectrum measured at 30~K for \textbf{H} $\bot$ $a_{h}$ fitted
by a Lorentzian line. Temperature dependence of (b) the ESR
intensity $I_{ESR}$ together with the \emph{dc} susceptibility
$\chi_{dc}$ measured at 3.4 kOe and (c) of the ESR linewidth. The
solid curve is a fit described in the text. Inset: sketch of the
spin relaxation via an Orbach process.}

\end{figure}

Fig.~\ref{Fig:ESR}(a) shows an ESR derivative absorption spectrum measured at
30 K. The spectrum is well fitted by a derivative
Lorentzian line shape characterized by effective $g$ factor,
peak-to-peak linewidth $\Delta H$, and double integrated intensity $I_{ESR}$.
The latter follows the temperature dependence
of the \emph{dc} susceptibility for $T<T^*$(=~120~K) (see
Fig.~\ref{Fig:ESR}(b)) as expected for an ESR signal originating
from the Cr$^{5+}$ dimers. Above $T^*$ the ESR absorption becomes
extremely broadened and cannot be tracked any further. In
Fig.~\ref{Fig:ESR}(c) we show the temperature dependence of the
linewidth $\Delta H$ with the magnetic field $H \perp a_h$.
The effective $g$-factor of 1.93(1) is almost constant below 70~K, and in good agreement with
THz transmission spectra and previous reports.\cite{Kofu09b,Aczel09}
When $\Delta H$ starts to increase strongly and reaches the order
of magnitude of the resonance field above 70 K (see below), $g$-factor
cannot be determined reliably anymore.

The temperature dependence is described using $\Delta H
= \Delta H_0 + A \exp{(-\frac{\Delta}{k_BT})}$ with a residual $\Delta H_0$ = 135~Oe, $A$ = 130~kOe, and a
gap $\Delta/k_B$ = 388~K. The exponential increase of the ESR
linewidth indicates that the relaxation of the excited spins occurs
via an Orbach process,\cite{Abragam1970} i.e. absorption of a
phonon with energy $\Delta$ to an excited orbital state and emission
of a phonon with energy $\Delta + E_Z$ where $E_Z$ is the Zeeman
splitting of the lower-lying $e$ orbital as illustrated in the inset
of Fig.~\ref{Fig:ESR}(c). Thus, the spin dynamics are
dominated by spin-lattice relaxation and we associate the gap
with the energy splitting of the Cr$^{5+}$ $e$-orbitals in the
regime $T<T^*$ where OO is complete but the system has not reached
its non-magnetic singlet ground state yet. The splitting $\Delta$
can be associated with the interaction driving OO and will be
important for a theoretical understanding of the energy scales in
\SrCrO.\cite{Radtke10} The fact that ESR spectra cannot
be observed above $T^*$ is ascribed to a drastically increased
spin-lattice relaxation rate for $T>T^*$ as a result of strong
fluctuations in the orbital and lattice DOF.
%Figure 2

%\subsection{Infrared active phonons: lattice fluctuations}\label{Sec:InfraredMode}

\begin{figure}[t]
\centering
\includegraphics[width=80mm,clip]{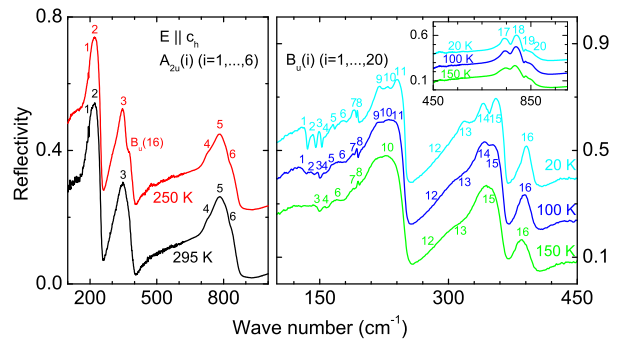}
\vspace{2mm} \caption[]{\label{Fig:sigma_prime_295K_20K} (Color
online)  Reflectivity spectra of \SrCrO~with $\mathbf{E}$ $\|$
$c_{h}$ (a) at 295~K and 250~K, and (b) at 150~K, 100~K and 20~K.
The spectra are shifted with respect to the ones at 150~K and 295~K in
order to clearly illustrate the evolution of phonon modes with
temperature.}

%In Figure (b), an additional mode around 330cm$^{-1}$ shows up, which locates at the same position as A$_{2u}$(3) in Figure (a). This leakage is observed probably due to the imperfect alignment.
\end{figure}

To study the lattice dynamics we performed IR reflectivity
measurements with the electric field \textbf{E}~$\parallel a_h$ and
\textbf{E}~$\parallel c_h$. Normal mode analysis yields the
irreducible representations of IR active modes $6A_{2u}
(\mathbf{E}\parallel z)+7E_{u}(\mathbf{E}\parallel (x,y))$ for the
hexagonal $R\overline{3}m$ (No.~166) structure.\cite{Chapon08,
Kroumova} The room temperature spectra for $\mathbf{E}\parallel
c_h$ (shown in Fig.~\ref{Fig:sigma_prime_295K_20K}(a)) and
$\mathbf{E}\parallel a_h$ (not shown) confirm this expectation by
exhibiting six modes with eigenfrequencies 192.8, 200.6, 335.6,
734.7, 765.6, and 827.3~cm$^{-1}$, and seven modes with
eigenfrequencies 110.6, 115.6, 183.0, 758.8, 805.0, 855.1, and
938.1~cm$^{-1}$, respectively. For the monoclinic $C2/c$ (No.~15b)
structure below JT transition, the number of expected normal
modes increases to $19A_{u} (\mathbf{E}\parallel
y)+20B_{u}(\mathbf{E}\parallel (x,z))$. However, only the selection
rule for the 19 $A_{u}$ modes is determined by $\mathbf{E}\parallel
b_m$, because the unique monoclinic $b_m$-axis coincides with a
principal axis of the dielectric tensor.  The dipole moments of the
20 $B_{u}$ modes are confined to the $a_m c_m$-plane, but their
directions do not necessarily coincide with the crystal axes. Using
the relations $a_h=\frac{1}{2}(a_m-b_m)$,
$b_h=-\frac{1}{2}(a_m+b_m)$, and $c_h=\frac{3}{2}c_m-\frac{1}{2}a_m$,\cite{Chapon08}
it is clear that upon cooling below $T_{JT}$ with
polarization $\mathbf{E}\parallel a_h$ we may expect to see at least
the 19 $A_{u}$ modes from the contribution of the monoclinic
$b_m$-axis plus possible additional $B_{u}$ modes due to the
projection of the polarization along $a_m$. Similarly, for
$\mathbf{E}\parallel c_h$ we can expect to probe the majority of the
20~$B_{u}$ modes confined to the $a_mc_m$-plane. The above
transformation corresponds to only one of three reported monoclinic
twins \cite{Castro10} and, thus, the number of expected normal modes
may be even larger. Consequently, one would expect a drastic
increase in the number of IR active modes for both measured
polarizations when comparing spectra below and above the JT
transition, e.g., at 250~K and 295~K. However, as shown in Fig.
\ref{Fig:sigma_prime_295K_20K}(a) the IR spectra do not
change dramatically across $T_{JT}$ = 285~K. Only one additional
mode $B_u(16)$ is already visible at 250~K, while the expected 20
$B_u$ modes of the low-temperature structure appear only gradually
upon further cooling. This behavior is illustrated in Fig.
\ref{Fig:sigma_prime_295K_20K}(b) where spectra at 150~K, 100~K, and
20~K are compared. The modes present at 20~K can be observed at
100~K, while in the phonon spectrum at 150~K several modes are not
resolved anymore and appear to be strongly broadened. A similar
behavior has also been observed for the IR spectra measured for
\textbf{E} $\|$ $a_{h}$. Previous structural diffraction studies
clearly assign the monoclinic symmetry to be realized just below
$T_{JT}$.\cite{Chapon08} Consequently, we interpret our results as
a signature of strong fluctuations which dominate the lattice
dynamics and lead to a strong damping and broadening of phonons in
the temperature range $T^*<T<T_{JT}$ with 100~K $<T^*<$ 150~K in
agreement with the ESR results.

%Figure 3

%\subsection{Specific heat: orbital fluctuations}\label{Sec:SpecHeat}

\begin{figure}[t]
\centering
\includegraphics[width=70mm,clip]{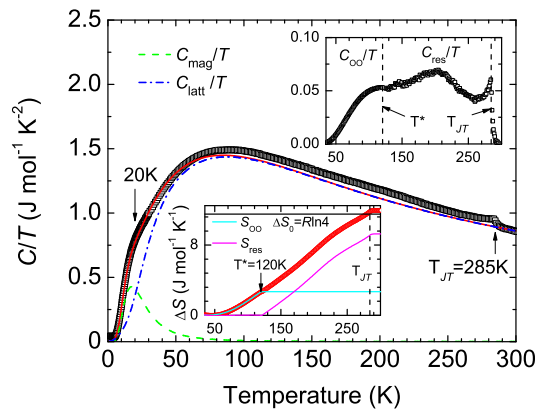}
\vspace{2mm} \caption[]{\label{Fig:Heat_capacity} (Color online)
Specific heat divided by temperature $C/T$ vs $T$.
Solid line is a superposition of the modeled magnetic (dashed line)
and phonon contribution (dash-dotted line) as described in the text.
Upper and lower insets show the temperature dependence of
orbital-related specific heat and entropy, respectively.}

%An anomaly of the specific heat can be observed around $T_{JT}$  285 K, which is zoomed in at the inset, is associated with the cooperative Jahn-Teller transition.
\end{figure}

Having identified the magnetic and phononic excitations we
turn to the specific heat (see
Fig.~\ref{Fig:Heat_capacity}). A peak without any thermal hysteresis,
which had not been detected in a previous specific-heat study,\cite{Singh07}
is clearly visible at $T_{JT}=285$ K, indicating an
order-disorder phase transition at 285~K. We associate
this anomaly with the OO transition reported to occur at 275~K.\cite{Chapon08}
Moreover, an additional broad shoulder is discernible around 20~K (see Fig.~\ref{Fig:Heat_capacity}).
We assume that the total heat capacity originates
from three different parts, a magnetic contribution
$C_{mag}$ corresponding to the thermal population of the excited
dimer states, a lattice contribution $C_{latt}$ due to phonons, and an electronic contribution reflecting the
orbital DOF. We approximate the magnetic contribution
by $C_{mag}(T)=N\frac{\partial E}{\partial T}$ using
$E=\frac{1}{Z}\sum_{i=0}^2 g_i\epsilon_i e^{-\beta \epsilon{_i}}$
with the partition function $Z=\sum_{i=0}^2 g_i e^{-\beta
\epsilon{_i}}$, the excitation energies
$\epsilon_{0,1,2}=0,h\nu_0^{opt},h\nu_0^{aco}$ as observed in the
THz transmission experiment, degeneracies $g_{0,1,2}=1,2,1$, and
$\beta\equiv 1/k_BT$. The resulting magnetic specific heat (dashed line in Fig. \ref{Fig:Heat_capacity})
accounts well for the shoulder at 20~K. Using the gap $\Delta$ determined by ESR we can fix the orbital
contribution $C_{oo}$ in the completely orbitally ordered phase
below $T^*$ by a two-level system with splitting $\Delta$ as shown
in the upper inset of Fig.~\ref{Fig:Heat_capacity}.

%\begin{equation}\label{eq:Capacity}
%C_{mag}(T)=R\beta^2\left[\frac{\sum_{i} E{_i}^2 g_i e^{(-\beta
%E{_i})}}{\sum_{i} g_i e^{(-\beta E_i)}}-\left(\frac{\sum_{i} E_i
%g{_i} e^{(-\beta E{_i})}}{\sum_{i} g_i e^{(-\beta
%E_i)}}\right)^2\right]
%\end{equation}
%where $R$ is the molar gas constant and $\beta\equiv 1/k_BT$ with
%Boltzmann constant $k_B$.  The characteristic energies are denoted
%by $E_{i}$'s with degeneracy $g_i$'s. This contribution is shown in
%Figure \ref{Fig:Heat_capacity} as a dashed line calculated from the
%energy gaps 1.24 THz $=59.5$ K and 1.47 THz $=70.6$ K, corresponding
%to the optical mode and acoustic mode, respectively, determined by
%THz transmission measurements discussed above.

%Figure 4

The lattice contribution can be described by a sum of one isotropic
Debye ($D$) and four isotropic Einstein terms ($E_{1,2,3,4}$). The
ratio between these terms was fixed to $D:E_1 :E_2
:E_3:E_4=1:3:4:3:2$ to account for the 39 DOF per
formula unit. The resulting contribution to the specific heat shown
as a dash-dotted line in Fig.~\ref{Fig:Heat_capacity} has been
obtained with the Debye and Einstein temperatures
$\theta_{D}=135.5$~K, $\theta_{E1}=153.4$~K, $\theta_{E2}=306.2$~K,
$\theta_{E3}=541.6$~K, and $\theta_{E4}=1360$~K consistent with
the frequency ranges where IR active phonons of the hexagonal
structure occur. Within these constraints the lattice contribution
was modeled in such a way that no discontinuity occurs between the
Schottky-like term for $T<T^*$ and the residual specific heat
$C_{res}$ = $C-C_{mag}-C_{latt}-C_{oo}$ for
$T>T^*$. In the upper inset of Fig.~\ref{Fig:Heat_capacity},
$C_{oo}$ and $C_{res}$ are shown together. One can clearly recognize
a $\lambda$-shaped anomaly at the JT transition at 285 K and a broad
hump-like contribution below the transition followed by the
Schottky-like contribution for $T<T^*$. The entropy $\Delta
S=S_{oo}+ S_{res}=\int_0^T
d\vartheta(C_{oo}+C_{res})/\vartheta$ associated with the orbital
DOF reaches a value slightly higher than the expected
$\Delta S_0=R\ln4$ (lower inset of Fig.~\ref{Fig:Heat_capacity}). We
interpret this observation as due to persistent fluctuations of the
orbital and lattice DOF in the temperature range
$T^*<T<T_{JT}$ in agreement with the anomalous temperature
dependence of the IR phonons and the ESR spectra.

In summary, the Zeeman splitting of singlet-triplet excitations at
5.13 and 6.08~meV was observed in \SrCrO. The spin-relaxation is dominated by
spin-lattice effects and revealed the splitting $\Delta/k_B$ = 388~K
of the low-lying $e$ doublet of the Cr ions below 120~K. The
broadening of spin resonances and polar phonons above 120~K is ascribed to strong orbital fluctuations. The specific
heat clearly marks the JT transition temperature at 285~K and reveals an extended fluctuation regime below the JT transition.
This indicates the competition of e.g. spin-orbit
coupling and electron-electron interactions with
electron-phonon coupling.

%{\color{red}In summary, we observed a Zeeman splitting of the excited dimer
%states and found the zero-field energies $6.08$ meV and $5.13$ meV
%of the cooperative acoustic and optical modes of the coupled dimer
%system.}

%We believe that a similar situation should occur to the related
%system \BaCrO~and we hope tohelps to understand the origin of BEC
%phenomena in systems with frustrated lattice geometries and strong
%electron-phonon coupling.

%\begin{acknowledgments}
We want to thank V. Tsurkan and D. Vieweg for experimental support and P. Lemmens, D. Wulferding, K.-H.
H\"{o}ck, R.M. Eremina, T.P. Gavrilova, and M.V. Eremin for
stimulating discussions. We acknowledge partial support by DFG via
TRR 80 and FOR 960.
%\end{acknowledgments}

%\begin{figure}[t]
%\centering
%\includegraphics[width=80mm,clip]{Figure4.eps}
%\vspace{2mm} \caption[]{\label{FCS_E2_lowTspec} (Color online) . }
%\end{figure}

%\bibliography{Gadolinium}

\begin{thebibliography}{39}

\bibitem{Tokura} Y. Tokura and N. Nagaosa, Science \textbf{288}, 462 (2000).
\bibitem{Khaliullin05} G. Khaliullin, Prog. Theor. Phys. Suppl. \textbf{160}, 155 (2005).
\bibitem{Krueger09} F. Kr\"{u}ger, S. Kumar, J. Zaanen, and J. van den Brink, Phys. Rev. B \textbf{79}, 054504 (2009).


\bibitem{Kugel82} K. I.~Kugel and D. I.~Khomskii, Sov.~Phys.~Usp.~{\bf 25}, 231
(1982).


%\bibitem{Saitoh01} E.~Saitoh \etal, Nature {\bf 410}, 180 (2001).

\bibitem{Saitoh01} E.~Saitoh, S. Okamoto, K. T. Takahashi, K. Tobe, K. Yamamoto, T. Kimura, S. Ishihara, S. Maekawa, and Y. Tokura, Nature {\bf 410}, 180 (2001).
%;R. Ruckamp \etal, New J.~Phys.~\textbf{7}, 144 (2005).

%\bibitem{Ruckamp05} R. Ruckamp \etal, New J.~Phys.~\textbf{7}, 144 (2005).



\bibitem{Khaliullin00} G. Khaliullin and S. Maekawa,  Phys. Rev. Lett. \textbf{85}, 3950 (2000);
V. Fritsch, J. Hemberger, N. B\"{u}ttgen, E.-W. Scheidt, H.-A. Krug von Nidda, A. Loidl, and V. Tsurkan, Phys. Rev. Lett.~\textbf{92}, 116401 (2004);
V. Tsurkan, O. Zaharko, F. Schrettle, Ch. Kant, J. Deisenhofer, H.-A. Krug von Nidda, V. Felea, P. Lemmens, J. R. Groza, D. V. Quach, F. Gozzo, and A. Loidl,
Phys. Rev. B \textbf{81}, 184426 (2010);
L. Balents, Nature \textbf{464}, 199 (2010).


%\bibitem{Fritsch04} V. Fritsch \etal, Phys. Rev. Lett. 92, 116401 (2004).

%\bibitem{Feiner05} L. F. Feiner and A. M. Oles, Phys. Rev. B 71, 144422 (2005).

%\bibitem{Balents10} L. Balents, Nature \textbf{464}, 199 (2010).


\bibitem{vandenBrink11} L. F. Feiner, A. M. Ole\'{s}, and J. Zaanen, Phys. Rev. Lett. \textbf{78}, 2799 (1997);
G. Jackeli and D. A. Ivanov, Phys. Rev. B, \textbf{76}, 132407 (2007).
%J. van den Brink \emph{et al}., in \emph{Introduction to Frustrated Magnetism}, edited by G. Lacroix \emph{et al}., (Springer, Heidelberg, 2011).
%Springer Series in Solid-State Sciences Vol. 164


%\bibitem{Zhou07} H. D. Zhou \emph{et al}., Phys. Rev. Lett. \textbf{99}, 136403 (2007).

%\bibitem{Tsurkan10} V. Tsurkan \emph{et al}., Phys. Rev. B 81, 184426 (2010).

\bibitem{Giamarchi08} T. Nikuni, M. Oshikawa, A. Oosawa, and H. Tanaka, Phys. Rev. Lett. \textbf{84}, 5868 (2000).
C. R\"{u}egg, N. Cavadini, A. Furrer, H.-U. G\"{u}del, K. Kr\"{a}mer, H. Mutka, A. Wildes, K. Habicht, and P. Vorderwisch, Nature (London) \textbf{423}, 62 (2003);
M. Jaime, V. F. Correa, N. Harrison, C. D. Batista, N. Kawashima, Y. Kazuma, G. A. Jorge, R. Stern, I. Heinmaa, S. A. Zvyagin, Y. Sasago, and K. Uchinokura, Phys. Rev. Lett. \textbf{93}, 087203 (2004);
T. Giamarchi, C. R\"{u}egg, and O. Tchernyshyov, Nature Physics \textbf{4}, 198 (2008).
%\bibitem{Nikuni}
%\bibitem{Jaime04}

%{\color{blue}\bibitem{Kofu09a}M. Kofu \emph{et al}., Phys. Rev. Lett. {\bf 102}, 037206 (2009).}

\bibitem{Aczel09} A. A. Aczel, Y. Kohama, C. Marcenat, F. Weickert, M. Jaime, O. E. Ayala-Valenzuela, R. D. McDonald, S. D. Selesnic, H. A. Dabkowska, and G. M. Luke, Phys.~Rev.~Lett. {\bf 103}, 207203 (2009).

\bibitem{Cuno89} E. Cuno and H. M\"{u}ller-Buschbaum, Z. Anorg. Allg. Chem. \textbf{572}, 95 (1989).

\bibitem{Chapon08} L. C. Chapon, C. Stock, P. G. Radaelli, and C. Martin, arXiv:0807.0877v2 (unpublished).


%\bibitem{Jahn} H. A. Jahn and E. Teller, Proc. Roy. Soc. A (London) \textbf{161}, 220 (1937).


\bibitem{Castro10} D. L. Quintero-Castro, B. Lake, E. M. Wheeler, A. T. M. N. Islam, T. Guidi, K. C. Rule, Z. Izaola, M. Russina, K. Kiefer, and Y. Skourski, Phys. Rev. B  \textbf{81}, 014415 (2010).


\bibitem{Radtke10} G. Radtke, A. Sa\'{u}l, H. A. Dabkowska, G. M. Luke, and G. A. Botton, Phys. Rev. Lett. \textbf{105}, 036401 (2010).

\bibitem{Islam10} A. T. M. Nazmul Islam, D. Quintero-Castro, Bella Lake, K. Siemensmeyer, K. Kiefer, Y. Skourski, and T. Herrmannsdorfer, Crystal Growth \& Design \textbf{10}, 465 (2010).

%\bibitem{Leuenberger}B. Leuenberger \emph{et al}., Phys. Rev. B {\bf 30}, 6300 (1984).


\bibitem{Kofu09b}M. Kofu, H. Ueda, H. Nojiri, Y. Oshima, T. Zenmoto, K. C. Rule, S. Gerischer, B. Lake, C. D. Batista, Y. Ueda, and S.-H. Lee, Phys. Rev. Lett. {\bf 102}, 177204 (2009).

%\bibitem{B.B.}B. Bleaney and K. D. Bowers, Proc. Roy. Soc. A (London) {\bf 214}, 451 (1952).

%\bibitem{Johnston}D. C. Johnston, in \emph{Handbook of Magnetic Materials}, Vol. 10, edited by K. H. J. Buschow (Elsevier Science, Netherlands, 1997), Page 1.

%\bibitem{footnote} The term $\cos(\frac{2}{3}n\pi)$ is used to write the
%eigenfrequencies in a compact way.


\bibitem{Glazkov04} V. N. Glazkov, A. I. Smirnov, H. Tanaka, and A. Oosawa \emph{et al}., Phys. Rev. B {\bf 69}, 184410 (2004); M. Matsumoto, T. Shoji, and M. Koga, J. Phys. Soc. Jpn. {\bf 77}, 074712 (2008).

\bibitem{Abragam1970} A. Abragam and B. Bleaney, {\it Electron
Paramagnetic Resonance of Transition Ions}, (Clarendon, Oxford,
1970).

\bibitem{Kroumova}E. Kroumova, M. I. Aroyo, J. M. Perez-Mato, A. Kirov, C. Capillas, S. Ivantchev, and H. Wondratschek, Phase Transitions, \textbf{76}, 155 (2003).

%\bibitem{Poulet1976} H. Poulet and J. P. Mathieu., {\it Vibration Spectra and Symmetry of Crystals}, Gordon and Breach, ??? (1976)
%\bibitem{Duarte}J. L. Duarte, J. A. Sanjurjo, andd R. S. Katiyar, Phys. Rev. B \textbf{36}, 3368 (1987).

%\bibitem{Kant}Ch. Kant, T. Rudolf, F. Schrettle, F. Mayr, J. Deisenhofer, P. Lunkenheimer, M. V. Eremin, and A. Loidl, Phys. Rev. B \textbf{78}, 245103, (2008).



\bibitem{Singh07} Y. Singh and D. C. Johnston, Phys. Rev. B \textbf{76}, 012407 (2007).

%\bibitem{Dzya}I. Dzyaloshinskii, J. Phys. Chem. Solids \textbf{4}, 241 (1958).

%\bibitem{Moriya}T. Moriya, Phys. Rev. \textbf{120}, 91 (1960).

%\bibitem{Room}T. R\~{o}\~{o}m, D. H\"{u}vonen, U. Nagel, Y.-J. Wang, and R. K. Kremer, Phys. Rev. B \textbf{69}, 144410 (2004).



\end{thebibliography}

%\clearpage

\end{document}